\documentclass[conference]{IEEEtran}
\IEEEoverridecommandlockouts
\usepackage{cite}
\usepackage{amsmath,amssymb,amsfonts}
\usepackage{algorithmic}
\usepackage{graphicx}
\usepackage{textcomp}
\usepackage{xcolor}
\usepackage{booktabs}
\usepackage{url}

\def\BibTeX{{\rm B\kern-.05em{\sc i\kern-.025em b}\kern-.08em
    T\kern-.1667em\lower.7ex\hbox{E}\kern-.125emX}}

\begin{document}

\title{Cross-Cutting Security Analysis of LLM-Generated Code via Metamorphic Testing and Association Rule Mining}

\author{
\IEEEauthorblockN{Zedong Peng}
\IEEEauthorblockA{\textit{University of Montana}\\
zedong.peng@umt.edu}
\and
\IEEEauthorblockN{Chenggang Wang}
\IEEEauthorblockA{\textit{University of Oklahoma}\\
chenggang.wang@ou.edu}
\and
\IEEEauthorblockN{Shangyue Zhu}
\IEEEauthorblockA{\textit{Central Washington University}\\
zhush@cwu.edu}
}

\maketitle

\begin{abstract}
Large language models (LLMs) frequently generate code with security vulnerabilities, yet these weaknesses are rarely isolated: they often span multiple concern areas simultaneously, reflecting the cross-cutting nature of security in software. We present a framework that combines security-oriented Metamorphic Relations (MRs) with Association Rule (AR) mining to detect vulnerabilities in LLM-generated code, uncover their co-violation structure, and trace that structure back to prompt-level risk factors. We define nine MRs covering major CWE categories, including SQL injection, XSS, command injection, path traversal, hard-coded credentials, weak cryptography, and memory-safety errors, and apply them using an LLM-based judge to 3,700 code snippets generated by five open models from the LLMSecEval benchmark. The results show that 68.8\% of snippets violate at least one MR, with hard-coded credentials (79.1\%) and command injection (74.4\%) among the most prevalent applicable failures. AR mining reveals strong cross-cutting co-violation patterns, notably that XSS and weak cryptography co-violations predict hard-coded credentials with 82.5\% confidence (lift = 3.23), along with tightly coupled clusters linking authentication, credential handling, and cryptographic weakness, as well as input-handling and memory-safety failures. We then perform prompt-level risk analysis and find that database- and authentication-related prompts are strong predictors of broad cross-cutting insecurity, while 65.5\% of prompts yield consistent violation outcomes across all five models. These findings show that insecure code generation is not merely a collection of independent defects, but a structured and prompt-conditioned phenomenon, motivating cluster-aware verification and prompt-level intervention for safer LLM-assisted programming.
\end{abstract}


\section{Introduction}

A developer asks an LLM to generate a simple login handler, a database query, or a file upload routine~\cite{wen2025ai}. The returned code may look functional at first glance, yet a single snippet can simultaneously omit authorization checks, embed hard-coded credentials, and mishandle untrusted input. In such cases, the problem is not one isolated bug, but a cross-cutting security failure that spans multiple concern areas at once.

This observation motivates our study. Prior work has shown that LLM-generated code often contains security weaknesses~\cite{pearce2022examining}, but most evaluations treat vulnerabilities as isolated categories, such as SQL injection, cross-site scripting, or buffer overflow, considered one at a time. In practice, however, security is a cross-cutting software property~\cite{kiczales1997aop}: failures in authentication, credential management, input validation, file handling, and memory safety may co-occur within the same generated artifact. Measuring only whether a snippet is vulnerable therefore misses an important question: how these vulnerabilities cluster together, and why certain prompts repeatedly induce broader security risk.

To address this gap, we present a framework that moves from detection, to structural diagnosis, to prompt-level explanation. We first define a catalog of nine security-oriented metamorphic relations (MRs) spanning major CWE categories and use an LLM-based judge to evaluate 3,700 code snippets generated from the LLMSecEval benchmark~\cite{tony2023llmseceval}. We then apply association-rule (AR) mining to the resulting violation matrix to uncover cross-cutting co-violation patterns, and finally link these patterns back to prompt characteristics through prompt-level risk analysis. This design builds on the MT-for-security lineage established by Rahman and Izurieta~\cite{rahman2023testing} for banking software and Chaleshtari et al.~\cite{chaleshtari2023metamorphic} for web systems, while extending it to LLM-generated code at scale and adding the AR mining and prompt analysis dimensions absent from prior work.

To investigate security as a cross-cutting property of LLM-generated code, we study three research questions:
\begin{list}{$\bullet$}{\leftmargin=1.2em \itemindent=0pt \topsep=2pt \parsep=0pt \itemsep=2pt}
    \item \textbf{RQ1:} How effectively can security-oriented metamorphic relations detect vulnerabilities in LLM-generated code across diverse CWE categories?
    \item \textbf{RQ2:} What cross-cutting co-violation patterns emerge from MR outcomes, and what do they reveal about the structure of security weaknesses in LLM-generated code?
    \item \textbf{RQ3:} Which prompt characteristics predict overall vulnerability burden and the activation of specific co-violation clusters across models?
\end{list}
Together, these questions move from detection to structural diagnosis to prompt-level explanation, enabling analysis of not only whether LLM-generated code is insecure, but how its vulnerabilities are organized and what prompt factors drive broader security risk.

This paper makes four contributions. First, we define a catalog of nine security-oriented metamorphic relations spanning major CWE categories and apply them to large-scale verification of LLM-generated code. Second, we combine MR-based analysis with association-rule mining to uncover cross-cutting co-violation patterns, showing that security failures form structured clusters rather than isolated defects. Third, we introduce a prompt-level risk analysis that links co-violation patterns to prompt characteristics, identifying which topics are associated with broader insecurity across models. Fourth, we evaluate the framework on 3,700 snippets from the LLMSecEval benchmark by five open models, providing empirical evidence that insecure code generation is highly prevalent and strongly conditioned by prompt content. All supplementary materials are available at \url{https://doi.org/10.5281/zenodo.19601101}.

\section{Background}

\textbf{Metamorphic Testing (MT).} A metamorphic relation $\text{MR} = \langle R_i, T, R_o \rangle$ specifies an input transformation $T$ and an expected output relation $R_o$; a violation occurs when $R_o$ does not hold~\cite{chen2018metamorphic}. For security testing, $T$ is an adversarial transformation (e.g., appending a SQL tautology) and $R_o$ encodes the expected secure behavior. MT is valuable when no test oracle exists, as with LLM-generated code where the correct secure implementation is not known a priori. Our MR judge performs \emph{static} analysis, reasoning about code structure rather than executing it.

\textbf{Association Rule Mining.} The Apriori algorithm~\cite{agrawal1994fast} extracts rules $X \Rightarrow Y$ from a transaction database, evaluated by \emph{support} (frequency), \emph{confidence} ($P(Y|X)$), and \emph{lift} ($P(Y|X)/P(Y)$, where lift $> 1$ indicates positive association). In our setting, each code snippet is a transaction and each violated MR is an item.

\textbf{Cross-Cutting Concerns.} Security is a canonical cross-cutting concern~\cite{kiczales1997aop}: input validation, authentication, and error handling cut across functional boundaries. We use AR mining to discover how security vulnerabilities co-occur across concern areas in LLM-generated code.

\section{Approach}

Fig.~\ref{fig:pipeline} summarizes the four-phase pipeline. We first generate code snippets from benchmark prompts (Phase~A), then judge each snippet against nine security-oriented MRs (Phase~B), mine the resulting violation matrix for co-violation patterns (Phase~C), and finally link those patterns back to prompt-level risk factors (Phase~D).

\begin{figure}[t]
\centering
\small
\setlength{\fboxsep}{4pt}
\fbox{\parbox{0.9\columnwidth}{
\textbf{Phase A:} 148 prompts $\times$ 5 models $\times$ 5 runs $\rightarrow$ 3,700 snippets\\[2pt]
\textbf{Phase B:} Each snippet $\xrightarrow{\text{9 MRs}}$ violation vector $\rightarrow$ matrix\\[2pt]
\textbf{Phase C:} Violation matrix $\xrightarrow{\text{Apriori}}$ association rules\\[2pt]
\textbf{Phase D:} AR patterns $\rightarrow$ prompt-level risk analysis
}}
\caption{Four-phase pipeline: generation, MR judgment, AR mining, and prompt-level risk analysis.}
\label{fig:pipeline}
\end{figure}

\subsection{MR Catalog}

Table~\ref{tab:mr_catalog} defines our nine MRs. Each MR specifies an adversarial transformation $T$ and the expected secure output relation $R_o$. Not all MRs apply to every snippet; applicability is determined per snippet based on its functionality.

\begin{table}[t]
\caption{Catalog of Nine Metamorphic Relations}
\label{tab:mr_catalog}
\centering
\small
\begin{tabular}{@{}llll@{}}
\toprule
\textbf{MR} & \textbf{CWE} & \textbf{Transform.\ $T$} & \textbf{Output $R_o$} \\
\midrule
MR$_1$ & 89 (SQLi) & Tautology inject. & $|f(x')| \leq |f(x)|$ \\
MR$_2$ & 79 (XSS) & Script tag inject. & HTML escaped \\
MR$_3$ & 78 (CmdInj) & Shell cmd append & No process spawn \\
MR$_4$ & 22 (PathTr) & Path trav.\ prefix & Access denied \\
MR$_5$ & 862 (Auth) & Unauth.\ user & Request rejected \\
MR$_6$ & 798 (Cred) & Static analysis & No hardcoded creds \\
MR$_7$ & 327 (Crypto) & Weak algo subst. & Rejected/warned \\
MR$_8$ & 120 (BOF) & Exceed buf.\ size & Bounds checked \\
MR$_9$ & 190 (IOF) & MAX\_INT + 1 & Overflow handled \\
\bottomrule
\end{tabular}
\end{table}

\subsection{MR Judge}

We use Claude Sonnet 4.6~\cite{anthropic2025claude} as an LLM-based judge to evaluate each snippet against all nine MRs simultaneously. The judge receives a structured prompt containing the code, its CWE context, the generating model name, and all nine MR definitions. For each MR, it returns: (1)~\emph{applicability}, whether the MR is relevant to the snippet's functionality (e.g., MR$_1$ is not applicable to code that performs no database operations); (2)~\emph{verdict}, violated, pass, or N/A; and (3)~\emph{evidence}, a one-sentence explanation supporting the verdict.

To ensure reproducibility, both the system prompt and user template are frozen at experiment initialization with SHA-256 hashes, and every judgment is saved individually with the full input prompt, raw LLM response, parsed verdicts, and wall-clock timing. A pilot run of 36~snippets was manually reviewed to validate template quality before the full run. All experimental artifacts (3,700 generation files, 3,700 judgment files, and an append-only audit log) are preserved for independent verification.

\subsection{AR Mining for Cross-Cutting Detection}

We construct a binary violation matrix $\mathbf{V} \in \{0, 1, \text{N/A}\}^{n \times 9}$ where $V_{ij} = 1$ indicates snippet $i$ violates MR$_j$, $V_{ij} = 0$ indicates a pass, and N/A entries are excluded from mining. We apply the Apriori algorithm~\cite{agrawal1994fast} with minimum support $\geq 0.01$ and minimum confidence $\geq 0.1$ to discover frequent co-violation itemsets and extract association rules. Rules with lift $> 1$ indicate positive co-occurrence beyond statistical independence, revealing cross-cutting security patterns. We additionally perform per-model AR mining to identify model-specific co-violation tendencies, and compute a $9 \times 9$ co-occurrence matrix to visualize pairwise MR relationships.

\subsection{Prompt-Level Risk Analysis}

For each of the 148 unique prompts, we extract features in four categories: (1)~\emph{structural features} (prompt length and word count); (2)~\emph{topic keywords} (database, web, authentication, cryptography, memory, and network); (3)~\emph{security awareness} (explicit mentions of ``secure,'' ``sanitize,'' or ``validate''); (4)~\emph{multi-topic complexity} (the count of distinct topic categories present). We correlate these features with the number of distinct MRs violated per prompt to identify high-risk characteristics. Additionally, we measure \emph{cross-model consistency}: for each prompt, we compute the fraction of models that produce at least one violation, revealing whether vulnerability is model-dependent or prompt-inherent.

\section{Evaluation}

\subsection{Experimental Setup}

\textbf{Dataset.} We use LLMSecEval~\cite{tony2023llmseceval}, comprising 150 natural-language prompts covering 18 CWEs in Python and~C. Two prompt identifiers appear twice in the dataset; after deduplication, 148 unique prompts remain. We use CodeQL as the automated baseline analyzer and LLMSecEval's secure code samples as reference implementations for sanity-checking MR applicability and judge validity.

\textbf{Models.} We focus on open models because they enable local execution, fixed-version evaluation, and stronger experimental reproducibility, all of which are important for security-sensitive analysis. We select models spanning three providers (Alibaba, DeepSeek, Google), three architecture types (Dense, MoE, PLE), and a parameter range from 4.5B to 30B to capture diversity in model design. Table~\ref{tab:models} lists the five models, all served locally via Ollama~\cite{ollama2024} with temperature 0.7 and max 1,024 tokens.

\begin{table}[t]
\caption{Code Generation Models}
\label{tab:models}
\centering
\small
\begin{tabular}{@{}llrl@{}}
\toprule
\textbf{Model} & \textbf{Provider} & \textbf{Params} & \textbf{Type} \\
\midrule
Qwen3-Coder 30B & Alibaba & 30B & MoE \\
Qwen2.5-Coder 7B & Alibaba & 7B & Dense \\
DeepSeek-Coder 6.7B & DeepSeek & 6.7B & Dense \\
CodeGemma 7B & Google & 7B & Dense \\
Gemma4:e4b & Google & 4.5B & PLE \\
\bottomrule
\end{tabular}
\end{table}

\textbf{Scale.} 148 unique prompts $\times$ 5 models $\times$ 5 runs = 3,700 code snippets (2,025 Python, 1,675~C), each judged against 9~MRs.

\textbf{Baselines.} We run four open SAST tools on all snippets: CodeQL v2.25.1~\cite{codeql2024} (taint analysis), Bandit 1.8.6~\cite{bandit2024} (Python pattern matching), Flawfinder 2.0.19~\cite{flawfinder2024} (C lexical analysis), and Semgrep 1.136.0~\cite{semgrep2024} (multi-language rule matching). Each tool's alerts are mapped to our nine MRs for direct comparison.

\subsection{RQ1: MR Effectiveness}

Table~\ref{tab:rq1_mr} reports per-MR detection results. Across all 3,700 snippets, \textbf{68.8\%} violate at least one MR. Hard-coded credentials (MR$_6$, 79.1\%) and command injection (MR$_3$, 74.4\%) show the highest violation rates among applicable snippets, while SQL injection (MR$_1$, 11.5\%) is lowest, suggesting LLMs partially internalize parameterized query patterns.

\begin{table}[t]
\caption{Per-MR Violation Rates (RQ1)}
\label{tab:rq1_mr}
\centering
\small
\begin{tabular}{@{}lrrr@{}}
\toprule
\textbf{MR} & \textbf{Applicable} & \textbf{Violated} & \textbf{Rate (\%)} \\
\midrule
MR$_1$ SQLi & 677 & 78 & 11.5 \\
MR$_2$ XSS & 1,234 & 468 & 37.9 \\
MR$_3$ CmdInj & 250 & 186 & 74.4 \\
MR$_4$ PathTrav & 670 & 471 & 70.3 \\
MR$_5$ AuthBypass & 848 & 536 & 63.2 \\
MR$_6$ HardCred & 823 & 651 & 79.1 \\
MR$_7$ CryptoWeak & 472 & 337 & 71.4 \\
MR$_8$ BuffOverflow & 1,647 & 785 & 47.7 \\
MR$_9$ IntOverflow & 1,083 & 528 & 48.8 \\
\midrule
\textbf{Any MR} & \textbf{3,700} & \textbf{2,545} & \textbf{68.8} \\
\bottomrule
\end{tabular}
\end{table}

Table~\ref{tab:rq1_model} compares models. DeepSeek-Coder (73.8\%) produces the most vulnerable code, while Gemma4 (65.1\%) is safest, a modest 8.7 percentage point spread across architectures. Notably, the largest model (Qwen3-Coder, 30B) does not produce the safest code; the smallest model (Gemma4, 4.5B) does, suggesting that model scale alone does not determine security quality in this parameter range. Across five independent runs per model-prompt pair, per-model violation rates vary with a standard deviation of 1.4 to 2.9 percentage points, confirming that the results are stable across stochastic generation.

\begin{table}[t]
\caption{Per-Model Violation Rates (RQ1)}
\label{tab:rq1_model}
\centering
\small
\begin{tabular}{@{}lrrr@{}}
\toprule
\textbf{Model} & \textbf{Total} & \textbf{Violated} & \textbf{Rate (\%)} \\
\midrule
DeepSeek-Coder 6.7B & 740 & 546 & 73.8 \\
CodeGemma 7B & 740 & 527 & 71.2 \\
Qwen2.5-Coder 7B & 740 & 496 & 67.0 \\
Qwen3-Coder 30B & 740 & 494 & 66.8 \\
Gemma4:e4b & 740 & 482 & 65.1 \\
\bottomrule
\end{tabular}
\end{table}

\textbf{SAST Baseline Comparison.} Table~\ref{tab:sast} compares four SAST tools with our MR approach. Individually, CodeQL detects 0.1\%, Bandit 1.2\%, Semgrep 2.4\%, and Flawfinder 30.5\%. Even combining all four, SAST detects 1,266 snippets (34.2\%), roughly half the MR approach's 68.8\%. The gap is largest for semantic MRs: no SAST tool detects any MR$_5$ (AuthBypass) violation, and only 44 of 651 MR$_6$ (HardCred) violations are caught. Flawfinder achieves strong coverage on MR$_8$ (BuffOverflow, 823 files) by pattern-matching dangerous C functions, but this lexical strategy does not generalize to Python or to MRs requiring behavioral reasoning.

\begin{table}[t]
\caption{SAST Baseline vs.\ MR Approach: Snippet-Level Detection}
\label{tab:sast}
\centering
\small
\begin{tabular}{@{}lrr@{}}
\toprule
\textbf{Tool} & \textbf{Detected} & \textbf{Rate (\%)} \\
\midrule
CodeQL 2.25.1 & 3 & 0.1 \\
Bandit 1.8.6 & 45 & 1.2 \\
Semgrep 1.136.0 & 90 & 2.4 \\
Flawfinder 2.0.19 & 1,130 & 30.5 \\
\midrule
Combined SAST (union) & 1,266 & 34.2 \\
\textbf{MR Approach} & \textbf{2,545} & \textbf{68.8} \\
\bottomrule
\end{tabular}
\end{table}

\subsection{RQ2: Cross-Cutting AR Patterns}

Apriori mining on the 2,545 violated snippets yields \textbf{44 association rules}. Table~\ref{tab:rq2_rules} presents the top rules by lift.

\begin{table}[t]
\caption{Top Association Rules by Lift (RQ2)}
\label{tab:rq2_rules}
\centering
\small
\begin{tabular}{@{}lcrr@{}}
\toprule
\textbf{Rule} & \textbf{Conf.} & \textbf{Lift} & \textbf{Supp.} \\
\midrule
XSS $\wedge$ Crypto $\Rightarrow$ HardCred & 0.825 & 3.23 & 0.013 \\
XSS $\wedge$ HardCred $\Rightarrow$ AuthByp & 0.592 & 2.81 & 0.024 \\
AuthByp $\wedge$ Crypto $\Rightarrow$ HardCred & 0.624 & 2.44 & 0.023 \\
XSS $\wedge$ HardCred $\Rightarrow$ Crypto & 0.320 & 2.42 & 0.013 \\
SQLi $\Rightarrow$ AuthBypass & 0.474 & 2.25 & 0.015 \\
Crypto $\Rightarrow$ HardCred & 0.570 & 2.23 & 0.075 \\
XSS $\wedge$ PathTrav $\Rightarrow$ BuffOvf & 0.630 & 2.04 & 0.011 \\
AuthByp $\wedge$ HardCred $\Rightarrow$ Crypto & 0.257 & 1.94 & 0.023 \\
AuthByp $\wedge$ BuffOvf $\Rightarrow$ XSS & 0.333 & 1.81 & 0.015 \\
HardCred $\Rightarrow$ AuthByp & 0.347 & 1.65 & 0.089 \\
\bottomrule
\end{tabular}
\end{table}

The expanded rule set in Table~\ref{tab:rq2_rules} reveals two cross-cutting \emph{clusters} and important inter-cluster bridges.

\textbf{Cluster 1: Authentication--Credential--Cryptography.}
The rules \{MR$_5$, MR$_6$, MR$_7$\} form a tightly coupled cluster, appearing in 9 of the top 10 rules. The rule Crypto $\Rightarrow$ HardCred (conf = 0.570, lift = 2.23, supp = 0.075) has the highest support among rules with lift $> 2$, indicating that 57\% of snippets using weak crypto also embed hard-coded secrets. This is a classic cross-cutting pattern: credential management and cryptographic concerns are logically independent, yet LLMs consistently fail on both simultaneously. The rule HardCred $\Rightarrow$ AuthByp (conf = 0.347, lift = 1.65, supp = 0.089) has the highest support of all 44 rules, indicating that hard-coded credentials and missing authorization checks are the most frequent co-violation pair overall, with 226 snippets exhibiting both.

The bidirectional nature of Cluster~1 is notable: XSS $\wedge$ HardCred $\Rightarrow$ Crypto (conf = 0.320, lift = 2.42) and AuthByp $\wedge$ HardCred $\Rightarrow$ Crypto (conf = 0.257, lift = 1.94) both predict weak cryptography when combined with HardCred, suggesting that HardCred acts as a \emph{hub vulnerability} that connects input validation failures to cryptographic weaknesses.

\textbf{Cluster 2: Input Handling--Memory Safety.}
XSS $\wedge$ PathTrav $\Rightarrow$ BuffOvf (conf = 0.630, lift = 2.04) connects web input handling with memory safety, and PathTrav--BuffOverflow co-occur in 170 snippets. Command injection (MR$_3$) and path traversal co-occur with lift = 2.34 in Qwen3-Coder, suggesting that models failing to sanitize file paths also fail to sanitize shell inputs.

\textbf{Inter-cluster bridge.} The rule AuthByp $\wedge$ BuffOvf $\Rightarrow$ XSS (conf = 0.333, lift = 1.81) bridges Clusters 1 and 2: when authentication bypass (Cluster~1) co-occurs with buffer overflow (Cluster~2), XSS is likely present, indicating that some prompts trigger failures spanning both clusters simultaneously.

\textbf{Per-model AR patterns.} Per-model mining reveals model-specific tendencies. DeepSeek-Coder produces the most rules (50) and the strongest SQLi--XSS--AuthBypass chain (MR$_1 \wedge$ MR$_2 \Rightarrow$ MR$_5$, conf = 1.0, lift = 3.82), indicating that when this model fails on input validation, it fails comprehensively. Gemma4:e4b produces the fewest rules (24), consistent with its lowest violation rate (65.1\%).

\textbf{Co-occurrence matrix.} The strongest off-diagonal pairs are MR$_5$--MR$_6$ (226 snippets), MR$_6$--MR$_7$ (192), MR$_4$--MR$_8$ (170), and MR$_2$--MR$_5$ (154). These map directly to the two clusters and confirm that security violations in LLM-generated code exhibit \emph{cross-cutting} structure: fixing one vulnerability class alone leaves correlated vulnerabilities unaddressed.

\subsection{RQ3: Prompt-Level Risk Analysis}

We classify 148 unique prompts into risk levels based on the number of distinct MRs violated across all models and runs: \textbf{high} ($\geq$4 MRs, 51 prompts, 34.5\%), \textbf{medium} (2--3 MRs, 68 prompts, 45.9\%), and \textbf{low} ($\leq$1 MR, 29 prompts, 19.6\%).

Table~\ref{tab:rq3_features} shows the prompt features most correlated with violation count. Database-related prompts are the strongest predictor ($r = 0.52$), followed by authentication topics ($r = 0.43$). Conversely, memory-focused prompts (array/buffer operations) show \emph{negative} correlation ($r = -0.37$), because these C-oriented prompts tend to trigger fewer diverse MR categories.

\begin{table*}[t]
\caption{Prompt Feature Analysis (RQ3). C1/C2: Cluster~1/2. H\%/L\%: prevalence in high-risk (51) vs.\ low-risk (29) prompts.}
\label{tab:rq3_features}
\centering
\small
\setlength{\tabcolsep}{4.5pt}
\begin{tabular}{@{}llrrrrrl@{}}
\toprule
\textbf{Feature} & \textbf{Keywords / Definition} & $r_\text{all}$ & $r_\text{C1}$ & $r_\text{C2}$ & \textbf{H\%} & \textbf{L\%} & \textbf{Interpretation} \\
\midrule
Database & sql, query, sqlite, mysql, db & +.52 & +.51 & $-$.04 & 49.0 & 3.4 & Strongest overall predictor; drives C1 \\
Authentication & password, login, auth, token, session & +.43 & +.54 & $-$.14 & 70.6 & 10.3 & Strongest C1 predictor \\
Multi-topic & Count of topic categories present & +.37 & +.27 & +.07 & 2.35$^\dagger$ & 1.14$^\dagger$ & More topic areas $\rightarrow$ more MRs \\
Memory & buffer, malloc, memcpy, array, pointer & $-$.37 & $-$.42 & +.03 & 3.9 & 31.0 & Anti-predicts C1; C-language prompts \\
Web & http, html, server, url, flask, django & +.21 & +.24 & $-$.07 & 41.2 & 10.3 & Moderate C1 association \\
Cryptography & encrypt, hash, md5, aes, cipher, ssl & +.19 & +.28 & $-$.04 & 13.7 & 0.0 & Predicts C1; absent in low-risk \\
File I/O & file, open, read, write, path, directory & +.06 & $-$.13 & +.31 & 33.3 & 41.4 & Only C2 predictor; hidden in $r_\text{all}$ \\
Security kw. & secure, sanitize, validate & $-$.04 & $-$.14 & +.12 & 0.0 & 3.4 & No effect on security \\
\bottomrule
\multicolumn{8}{l}{\footnotesize $^\dagger$Mean topic count, not prevalence \%.}
\end{tabular}
\end{table*}

Table~\ref{tab:rq3_features} reveals that the two clusters respond to distinct prompt characteristics. Database and authentication keywords strongly predict Cluster~1 ($r_\text{C1} = +0.51$ and $+0.54$) but not Cluster~2, while file I/O keywords are the sole Cluster~2 predictor ($r_\text{C2} = +0.31$) yet appear invisible in the overall correlation ($r_\text{all} = +0.06$). This shows that per-cluster analysis uncovers associations that total-count analysis misses. Memory keywords anti-predict Cluster~1 ($r_\text{C1} = -0.42$) because C-oriented prompts trigger narrow MR categories. Security-aware phrasing shows no effect ($r_\text{all} = -0.04$, present in 0\% of high-risk prompts), indicating that explicitly requesting secure code does not improve output safety.

\textbf{CWE vulnerability profiles.} Vulnerability patterns vary substantially across CWEs. CWE-434 (Unrestricted Upload) prompts trigger an average of 5.2 distinct MRs per prompt, the highest of any CWE, because file upload functionality involves authentication, path handling, and input validation simultaneously. In contrast, CWE-190 (Integer Overflow) prompts trigger only 0.9 MRs on average, as overflow checking is a narrow, isolated concern. CWE-89 (SQL Injection) prompts average 4.3 MRs, suggesting that database-facing code is inherently multi-concern.

\textbf{Cluster-specific prompt features.} To connect prompt-level risk factors with the co-violation structure identified by AR mining, we analyze which prompt features predict each cluster rather than only the total number of violated MRs. The two clusters respond to distinct prompt characteristics. Cluster~1 (\{AuthBypass, HardCred, CryptoWeak\}) is strongly predicted by authentication keywords ($r = 0.54$) and database keywords ($r = 0.51$), while memory-related prompts show a negative association ($r = -0.42$). Cluster~2 (\{CmdInj, PathTrav, BuffOverflow\}) is primarily predicted by file I/O keywords ($r = 0.31$) and shows weak dependence on other features, indicating that memory safety violations arise broadly rather than from a specific prompt topic.

Of the 148 prompts, 51 (34.5\%) trigger both clusters, 20 (13.5\%) trigger only Cluster~1, 63 (42.6\%) trigger only Cluster~2, and 14 (9.5\%) trigger neither. Among high-risk prompts, 78.4\% activate both clusters simultaneously, compared to 16.2\% of medium-risk and 0\% of low-risk prompts. This indicates that \emph{cross-cluster} activation, not just high violation count, is the hallmark of prompt-level risk. Multi-topic prompts (spanning two or more keyword categories) trigger both clusters at 44.4\%, compared to 14.3\% for single-topic prompts, confirming that the cross-cutting nature of security violations is driven by the multi-concern nature of the prompt itself.

\textbf{Cross-model consistency.} In 65.5\% of prompts, all five models agree on violation status. Of the 148 prompts, 94 (63.5\%) produce violations across \emph{all} models, while only 3 (2.0\%) are consistently secure. The remaining 51 (34.5\%) show mixed results where some models succeed and others fail. The mean cross-model agreement score is 0.846 (where 1.0 indicates all models produce the same outcome). This high agreement indicates that prompt-level vulnerability is largely model-agnostic; the prompt itself, not the model architecture, is the primary driver of insecure code.

The 51 mixed-result prompts are the most actionable for practitioners: they represent cases where model selection or ensemble strategies could improve security outcomes. Among these mixed prompts, Qwen3-Coder~30B is the least likely to violate (49.0\%), while DeepSeek-Coder is most likely (76.5\%). Notably, the ranking on mixed prompts differs from overall rates: Qwen3-Coder outperforms Gemma4:e4b (62.7\%) on contested prompts despite similar overall rates (66.8\% vs.\ 65.1\%), suggesting that the larger model's advantage emerges specifically on ambiguous security scenarios.

\subsection{Discussion}

Our results carry three practical implications. First, the AR-discovered cross-cutting clusters suggest a \emph{prioritization strategy}: when one MR violation is detected, developers should proactively check the associated MRs from the same cluster. For instance, detecting a hard-coded credential (MR$_6$) should trigger immediate review for weak cryptography (MR$_7$, 57\% conditional probability) and authentication bypass (MR$_5$, 34.7\%). This cluster-based prioritization could reduce manual review effort by focusing attention on the most likely co-occurring vulnerabilities.

Second, the cluster-specific prompt analysis (connecting RQ2 and RQ3) enables \emph{targeted} prompt-level intervention. Since database and authentication keywords specifically predict Cluster~1 violations while file I/O keywords predict Cluster~2, automated tools can flag prompts by expected violation type, not just overall risk. For example, a prompt mentioning both database operations and file handling would be flagged for both clusters, warranting comprehensive security review. The negligible correlation between security-aware phrasing and secure output ($r = -0.04$) reinforces that current LLMs do not respond to surface-level security instructions, making post-generation verification essential.

Third, the high cross-model consistency (65.5\% agreement, mean = 0.846) indicates that vulnerability is predominantly prompt-driven rather than model-dependent. This has implications for mitigation strategies: improving the prompt or the training data may be more effective than switching between models of comparable size.

\subsection{Threats to Validity}

\textbf{Internal.} Our MR judge uses an LLM (Claude Sonnet 4.6) rather than runtime execution. While pilot validation of 36 judgments showed high accuracy, LLM-based judges may exhibit systematic biases. The frozen prompt templates and per-judgment provenance data support auditability. To mitigate judge variance, we generate five runs per model--prompt pair and aggregate results.

\textbf{External.} Results are specific to LLMSecEval prompts and five open models (4.5B--30B). Generalization to production codebases, larger models, or closed models (e.g., GPT-4, Claude) requires further study. AR patterns may differ for datasets with different CWE distributions or languages beyond Python and~C.

\textbf{Construct.} Our nine MRs cover major CWE categories but do not exhaust all security concerns (e.g., race conditions, information leakage via CWE-200 side channels). The binary violated/pass classification may oversimplify nuanced security states. Additionally, our static MR formulation, while scalable, may miss vulnerabilities that only manifest at runtime.

\section{Related Work}

Metamorphic testing has been applied to software security by Rahman and Izurieta~\cite{rahman2023testing} for banking systems and Chaleshtari et al.~\cite{chaleshtari2023metamorphic} for web applications, building on the MT framework surveyed by Chen et al.~\cite{chen2018metamorphic}. Separately, LLMORPH~\cite{cho2025llmorph} and METAL~\cite{hyun2024metal} apply MT to LLM behavior testing, but focus on functional correctness and fairness rather than security. Our work extends the MT-for-security lineage to LLM-generated code at scale and adds AR mining and prompt-level risk analysis as new diagnostic dimensions.

Several studies measure vulnerability rates in LLM-generated code. Pearce et al.~\cite{pearce2022examining} found that Copilot produces vulnerable code in approximately 40\% of security-relevant scenarios; benchmarks such as LLMSecEval~\cite{tony2023llmseceval}, CyberSecEval3~\cite{bhatt2023cyberseceval}, and Cweval~\cite{peng2025cweval} provide evaluation datasets. Shen et al.~\cite{shen2026failure} studied failure-aware decision frameworks. On the generation side, Patir et al.~\cite{patir2025grasp} use graph-based reasoning to fortify LLM code generation, and LLMSecCode~\cite{shahid2025llmcsec} evaluates secure coding practices. Unlike these works, our framework moves beyond measuring vulnerability rates to diagnosing the cross-cutting co-violation structure and linking it to prompt-level risk factors.

\section{Conclusion}

We presented a framework combining Metamorphic Relations with Association Rule mining for cross-cutting security verification of LLM-generated code. Our evaluation on 3,700 code snippets generated by five open models from the LLMSecEval benchmark yields several key findings:

\begin{list}{$\bullet$}{\leftmargin=1.2em \itemindent=0pt \topsep=2pt \parsep=0pt \itemsep=2pt}
    \item \textbf{High vulnerability prevalence:} 68.8\% of snippets violate at least one security MR, with hard-coded credentials (79.1\%) and command injection (74.4\%) being the most common.
    \item \textbf{Cross-cutting structure:} AR mining yields 44 rules, 29 with lift $> 1$, revealing two principal clusters: \{AuthBypass, HardCred, CryptoWeak\} and \{PathTrav, BuffOverflow, CmdInj\}.
    \item \textbf{Prompt-driven vulnerability:} High-risk prompts containing database keywords are 14.2 times more likely, and those with authentication keywords 6.8 times more likely, to trigger cross-cutting violations than low-risk prompts. In 65.5\% of prompts, all five models produce consistent results.
\end{list}

These results demonstrate that security in LLM-generated code is not merely a collection of independent vulnerabilities but a structured, cross-cutting phenomenon amenable to systematic diagnosis through data mining. The cross-cutting clusters discovered by AR mining, particularly the \{AuthBypass, HardCred, CryptoWeak\} triad, suggest that LLMs lack a holistic model of security: they may learn to parameterize SQL queries (MR$_1$ violation rate of only 11.5\%) without learning to avoid hard-coding the database credentials in the same snippet (MR$_6$ at 79.1\%).

Future work includes: (1)~extending to runtime-executable MR checking for higher fidelity verdicts; (2)~using AR rules as feedback to guide prompt rewriting for safer code generation, creating a closed-loop security improvement system; and (3)~developing IDE plugins that automatically flag high-risk prompt patterns identified by our prompt-level risk analysis. All experimental artifacts, including 3,700 generation files, 3,700 judgment files with full provenance, and analysis outputs, are preserved for reproducibility.

\bibliographystyle{IEEEtran}
\bibliography{references}

\end{document}